# Seeing full vectorial structures of light fields with a single-shot holographic multiplexed detector

*Sitao Qin[+,1], Yize Liang[+,1,2]\*, Shuai Cao[1], Changqing Cao[1], Xukun Yin[1], Lixian Liu[1], Mingjian Cheng[3], and Huailiang Xu[1,4]\**

[1] School of Optoelectronic Engineering, Xidian University, Xi'an 710071, China

[2] Key Laboratory of Light Field Manipulation and System Integration Applications in Fujian Province, School of Physics and Information Engineering, Minnan Normal University, Zhangzhou 363000, China

[3] School of physics, Xidian University, Xi'an 710071, China

[4] State Key Laboratory of Integrated Optoelectronics, College of Electronic Science and Engineering, Jilin University, Changchun 130012, China

[+]These authors contributed equally to this work.

[*]E-mail: liangyize@xidian.edu.cn, xuhuailiang@xidian.edu.cn

**Abstract**

**The vectorial structure of light—amplitude, phase, and polarization—encodes essential information for applications ranging from super-resolution microscopy to high-capacity communications and quantum information processing. However, existing characterization methods either rely on multiple sequential measurements or require bulky polarization-splitting optics in the signal path. Here we propose and experimentally demonstrate a single-shot holographic multiplexed detector that retrieves the full vectorial information from a single intensity recording. Two orthogonally polarized reference beams with distinct off-axis carriers interfere with the unknown vectorial light field, encoding both polarization channels into one off-axis hologram. Digital holographic reconstruction combined with a self-calibrated global phase retrieval recovers the complex wavefronts in the two channels without any additional measurements. We validate our approach by characterizing the polarization structures and concurrence of various vectorial structured light beams on a higher-order Poincaré sphere ($l$=2). This compact, efficient detector may open new routes for real-time vectorial metrology in light–matter interaction, chiral sensing, vectorial adaptive optics, and dynamic structured light applications.**

**Introduction**

Polarization is the vectorial signature of light, as fundamental as its intensity or wavelength yet encoding distinct and invisible structural information[1,2]. From ellipsometric thin-film characterization and remote sensing of aerosol particles to secure quantum key distribution, this vectorial dimension is indispensable wherever light interacts with matter or carries information[3-7]. Recently, manipulation of the polarization of light beams has evolved from spatially homogeneous polarized light to arbitrary vectorial light fields[8-13]. Such vectorial fields have become enabling tools in super-resolution microscopy[14], optical trapping[15,16], optical communications[17,18], and structured light quantum information processing[19,20]. Dealing with them, however, involves more complex scenarios. This increasing complexity, in turn, places higher demands on the techniques used to detect and characterize the full vectorial information of such light fields, including polarization structures, vectorness/concurrence, and polarization topology, to name a few[21-24].

To characterize the polarization state of light, two vectorial formalisms have been established: the Stokes vector and the Jones vector[25]. The Stokes vector consists of four real parameters that describe the polarization state in terms of measurable intensities[26,27]. The Jones vector, in contrast, represents fully polarized light with two complex amplitudes, encoding the relative amplitude and phase between orthogonal polarization components[28,29]. Correspondingly, measuring either the Stokes or Jones vector at each point across the transverse plane defines two major approaches for spatially resolved vectorial light field characterization. The former requires Stokes polarimetry[17,30-33], i.e., four intensity measurements for the Stokes parameters, whereas the latter requires measuring the complex wavefronts of the beam in two orthogonal uniformly, polarized channels[34-36]. Both methods are powerful, yet their complexity motivates the development of a truly single-shot approach that can efficiently retrieve the full vectorial information.

In this work, we propose and experimentally demonstrate a truly single-shot holographic multiplexed detector for efficient vectorial structure measurement. Two reference beams, uniformly polarized in orthogonal channels, are directed to off-axis interfere with the unknown vectorial light field. By introducing distinct off-axis angles for the two orthogonally polarized reference beams, we achieve multiplexing of two off-axis interferences in a single hologram. Performing off-axis digital holography on the recorded multiplexed hologram enables efficient retrieval of the complex wavefronts in the two orthogonally polarized channels. Considering the optical path difference between the two reference beams, a global phase calibration is performed on the complex wavefronts of the two channels to recover the full vector information. Using the proposed detector, we characterize the full vectorial structures of various vectorial structured light beams on a higher-order Poincaré sphere (HOPS). This approach offers new insights for vectorial information measurement and is expected to empower a broad class of light–matter interaction studies.

**Concept**

To begin with, we recall the concept of traditional vectorial light detection methods based on Jones matrix characterization. Such an approach requires measuring the complex wavefronts of the tested vectorial structured light in two orthogonal polarization channels, e.g., *H*-(horizontal) and *V*-(vertical) polarization channels. Without loss of generality, off-axis digital holography can be employed to measure the complex wavefronts in these two orthogonal channels. Consider an unknown signal beam $E_s(x,y) = |E_s(x,y)|e^{i\varphi_s(x,y)}$ that interferes off-axis with a reference Gaussian beam $E_R(x,y) \approx |E_R|e^{i(\varphi_R+2\pi u_0 x)}$. The interference intensity can be expressed as,

$$|E_s(x,y)+E_R(x,y)|^2 = |E_s(x,y)|^2 + |E_R|^2 + |E_s(x,y)||E_R|e^{i[\varphi_s(x,y)-\varphi_R]}e^{-i2\pi u_0 x} + \{|E_s(x,y)||E_R|e^{i[\varphi_s(x,y)-\varphi_R]}e^{-i2\pi u_0 x}\}^*, \quad (1)$$

where $2\pi u_0 x$ denotes the off-axis carrier in the $x$-direction, and * is complex conjugation. In off-axis digital holography, a large Gaussian beam (much larger than the tested beam) is used, such that its intensity distribution and phasefront can be approximated by $|E_R|$ and global phase term $e^{i\varphi_R}$, respectively. Performing a 2D fast Fourier transform (FFT) on the interference intensity, the cross-correlation (CC) term becomes separated from the direct-current (DC) term due to the Fourier shift factor $e^{-i2\pi u_0 x}$. This FFT process can be expressed as[37],

$$\mathcal{F}\{|E_s(x,y)+E_R(x,y)|^2\} = DC(u,v) + CC(u+u_0,v) + CC(u-u_0,v)^*, \quad (2)$$

where DC term is the FFT of $|E_s(x,y)|^2 + |E_R|^2$, CC term is the FFT of $|E_s(x,y)||E_R|e^{i[\varphi_s(x,y)-\varphi_R]}$. Thus, by filtering out the CC($u-u_0,v$) term in the spatial-frequency (SF) domain, shifting it to the center, and performing inverse fast Fourier transform (FFT), one can rapidly acquire the complex wavefront term $|E_s(x,y)||E_R|e^{i[\varphi_s(x,y)-\varphi_R]}$.

As illustrated in Fig. 1a and b, after splitting the vectorial structured light beam into two orthogonal polarization components, off-axis digital holography can be performed in both channels to retrieve their complex wavefronts. The polarization structure of the tested vectorial structured light beam can then be determined from these measured complex wavefronts. It is worth noting that a global phase calibration process is required because of the optical path difference between the two reference beams. Once the polarization structure is obtained, a wealth of vectorial information can be characterized, including, for example, concurrence, vectorness, non-separability, or the polarization topology of skymirons. While this traditional approach is effective for characterizing vectorial light fields, it still suffers from several drawbacks, such as requiring at least two measurements and introducing polarization beam-splitting devices into the signal path.

To overcome these drawbacks, we propose and experimentally demonstrate a truly single-shot holographic multiplexed detector for efficient vectorial structure measurement, as displayed in Fig. 1 c. By providing different off-axis carriers for the two $H$-and $V$-polarized reference Gaussian beams and allowing them to interfere with a vectorial structured light beam, two off-axis interference holograms can be multiplexed into a single intensity profile. The tested vectorial structured light beam and the off-axis reference beams can be respectively expressed as,

$$\begin{cases} E_{S,H}(x,y) = |E_{S,H}(x,y)|e^{i\varphi_{S,H}(x,y)} \\ E_{S,V}(x,y) = |E_{S,V}(x,y)|e^{i\varphi_{S,V}(x,y)} \end{cases}, \quad \begin{cases} E_{R,H}(x,y) \approx |E_{R,H}|e^{i(\varphi_{R,H}+2\pi u_0 x)} \\ E_{R,V}(x,y) \approx |E_{R,V}|e^{i(\varphi_{R,V}+2\pi v_0 y)} \end{cases}, \quad (3)$$

where $|E_{S,H}(x,y)|$ and $\varphi_{S,H}(x,y)$ are the amplitude and phasefront terms of the $H$-polarized component of the tested vectorial structured light beam, $|E_{S,V}(x,y)|$ and $\varphi_{S,V}(x,y)$ are the amplitude and phasefront terms of the $V$-polarized component of tested vectorial structured light beam, $\varphi_{R,H}$ and $\varphi_{R,V}$ are global phase terms of the $H$-polarized and $V$-polarized reference beams respectively, $2\pi u_0 x$ and $2\pi u_0 y$ are different off-axis carriers towards the $x$ and $y$ directions. The multiplexed interference intensity profile can be written as,

$$I_{Interference}(x,y) = |E_{S,H}(x,y)+E_{R,H}(x,y)|^2+|E_{S,V}(x,y)+E_{R,V}(x,y)|^2$$
$$= |E_{S,H}(x,y)|^2+|E_{R,H}|^2+|E_{S,V}(x,y)|^2+|E_{R,V}|^2+cc_1+cc_1^*+cc_2+cc_2^*, \quad (4)$$

where $|E_{S,H}(x,y)|^2+|E_{R,H}|^2+|E_{S,V}(x,y)|^2+|E_{R,V}|^2$ is the direct current term, $cc_1$ and $cc_2$ are CC terms which can be expressed as $cc_1 = |E_{S,H}(x,y)||E_{R,H}|e^{i[\varphi_{S,H}(x,y)-\varphi_{R,H}]}e^{-i2\pi u_0 x}$ and $cc_2 = |E_{S,V}(x,y)||E_{R,V}|e^{i[\varphi_{S,V}(x,y)-\varphi_{R,V}]}e^{-i2\pi v_0 y}$. Performing 2D FFT on $I_{Interference}(x,y)$, the corresponding SF domain can be written as,

$$\mathcal{F}\{I_{Interference}(x,y)\}=DC(u,v)+CC_1(u+u_0,v)+CC_1(u-u_0,v)^*+CC_2(u,v+v_0)+CC_2(u,v-v_0)^*, \quad (5)$$

the $CC_1$ and $CC_2$ terms are separated due to their distinct Fourier shift factors $e^{-i2\pi u_0 x}$ and $e^{-i2\pi v_0 y}$. Thus, performing a 2D IFFT on $CC_1(u,v)$ and $CC_2(u,v)$ enables rapid acquisition of the terms $|E_{S,H}(x,y)||E_{R,H}|e^{i[\varphi_{S,H}(x,y)-\varphi_{R,H}]}$ and $|E_{S,V}(x,y)||E_{R,V}|e^{i[\varphi_{S,V}(x,y)-\varphi_{R,V}]}$ in a single-shot process, as indicated in Fig. 1d. Since large Gaussian beams serve as reference beams, $|E_{R,H}|$, $|E_{R,V}|$, $\varphi_{R,H}$, and $\varphi_{R,V}$ can be approximately treated as constants. Ensuring that the two reference beams have the same power (i.e., $|E_{R,H}| = |E_{R,V}| =$ constant), these amplitude terms do not affect the final retrieved polarization structures. However, because the two reference beams propagate along distinct optical paths, their different global phase terms $\varphi_{R,H}$ and $\varphi_{R,V}$ introduce measurement errors in the final polarization structures. Therefore, a global phase calibration process is required to accurately characterize the polarization structure of the measured vectorial structured light beam.

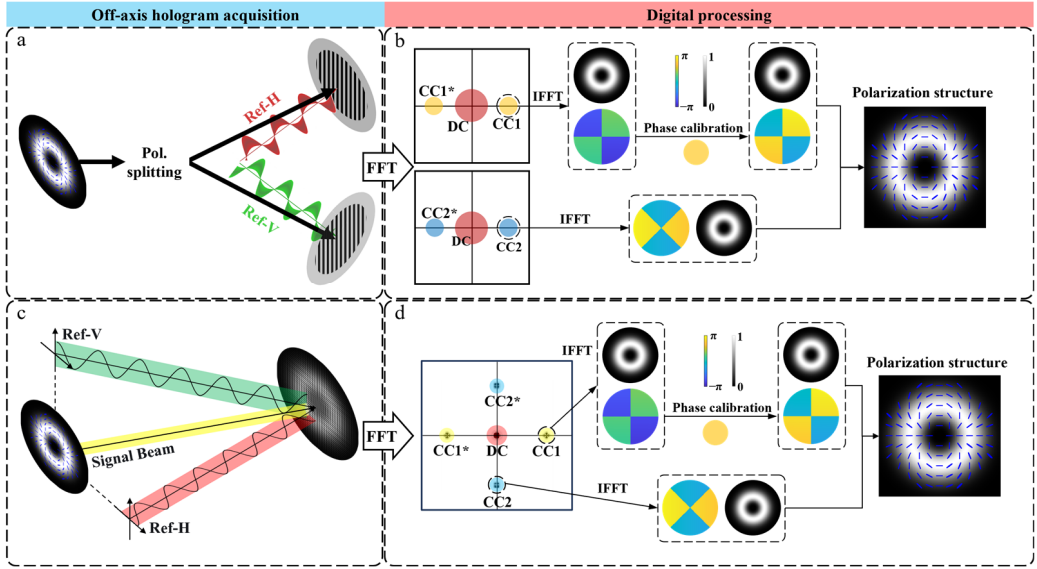

**Fig 1 | A conceptual comparison between the traditional holographic vectorial light field detector and the proposed holographic multiplexed detector. a,** Off-axis hologram acquisition process for the traditional detector. The tested vectorial structured light beam is polarization-split into H- and V- polarized components, which then respectively interfere off-axis with two reference beams. **b,** Digital processing principle for the traditional detector, including performing FFT to obtain the spatial frequency (SF) domain, acquiring complex wavefronts via IFFT, and applying global phase calibration to retrieve the final polarization structure. **c** Off-axis hologram acquisition process for our proposed detector. The tested vectorial structured light beam directly interferes off-axis with two reference beams. **d** Digital processing principle for the proposed detector, including performing FFT to obtain the SF domain, acquiring complex wavefronts via IFFT, and applying global phase calibration to obtain the final polarization structure.

In this work, we propose a self-calibrated method to determine the global phase difference $\varphi_{R,H}-\varphi_{R,V}$ for calibration. For the vectorial structured light beam shown in Fig. 1c, because the integrals of $\varphi_{S,H}(x,y)$ and $\varphi_{S,V}(x,y)$ over the entire plane are equal, one can calculate the difference between the mean values of the measured phasefronts in *H*- and *V*- polarization channels to obtain $\varphi_{R,H}-\varphi_{R,V}$ (see Supplementary Note 2 for details). Such a digital self-calibrated approach does not require any additional measurement, thereby ensuring the efficiency of our proposed detector.

**Simulation and experimental results**

To test the efficiency of the proposed holographic multiplexed detector, we first characterize the polarization structure of a specific vector beam with a modal index of 2 in both simulation and experiment. Such a vector beam can be generated by a linear superposition of two optical vortex (OV) beams with opposite orbital angular momenta (OAM) and spin angular momenta (SAM), expressed as $OV_{+2}\hat{R}+OV_{-2}\hat{L}$, where $OV_{+2}$ and $OV_{+2}$ are OV beams with topological charges (TCs) of +2 and -2 respectively, and $\hat{R}$ and $\hat{L}$ correspond to right and left circular polarizations.

Figures 2a to 2g illustrate the simulation results for characterizing the polarization structure of such a vector beam. Shown in Fig. 2a is the simulated off-axis hologram for multiplexed interference. Since the two reference beams possess orthogonal off-axis carriers in the *x* and *y* directions, linear fringes along two different directions can be observed in the off-axis hologram, particularly in the enlarged view. Performing a FFT on this hologram yields the SF domain shown in Fig. 2b. In the SF domain, the $CC_1$, $CC_2$, and DC terms are completely separated. Therefore, by extracting the $CC_1$ and $CC_2$ terms, shifting them to the center of the SF domain, and performing an IFFT on them, the complex wavefronts of the *H*- and *V*- polarized components can be acquired. Correspondingly, Figs. 2c and 2d present the retrieved complex wavefront of the *H*- polarized component, while Figs. 2e and 2f show that of the *V*- polarized component. Once the complex wavefronts of the *H*- and *V*- polarized components are known, a global phase calibration process enables obtaining the accurate Jones vector of each point on the transverse plane. Finally, plotting polarization ellipses on the intensity profile[38] (i.e., the sum of the intensity profiles in Figs. 2c and 2e) of the vector beam allows one to determine the polarization structure, as shown in Fig. 2g.

Correspondingly, Figs. 2h–2n present the experimental results for characterizing the polarization structure of the vector beam. Figure 2h shows the captured off-axis hologram, with its enlarged view displayed on the right. It can be observed that the contrast of the interference fringes decreases to some extent in the experiment, which may be caused by mechanical jitter during the single exposure time. Although this reduction in interference contrast affects the signal-to-noise ratio (SNR), the SNR remains sufficiently high to extract the polarization information of the vector beam. Figure 2i shows the corresponding SF domain, where the CC terms are well separated from the DC term. The retrieved intensity profiles and phasefronts of the *H*- and *V*- polarized channels are displayed in Figs. 2j–2m. Finally, applying a global phase calibration process enables determination of the experimentally measured polarization structure, as illustrated in Fig. 2n, which is in good agreement with the simulation result shown in Fig. 2g.

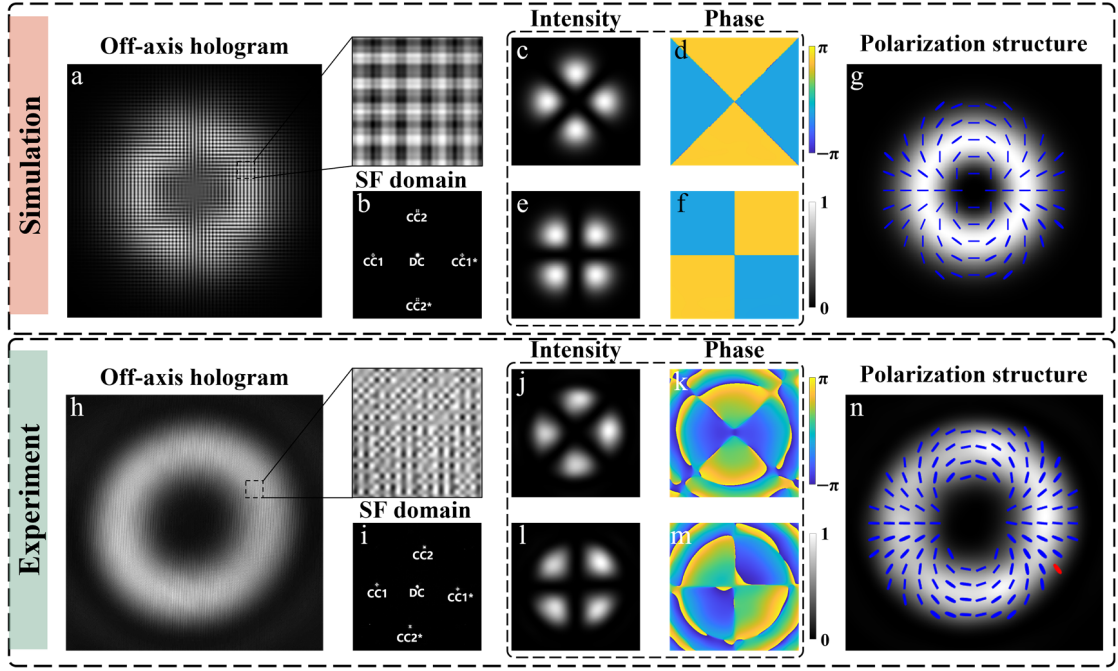

**Fig 2 | Simulation and experimental results for detecting the polarization structure of a vector beam with modal index of 2.** **a-g,** Simulation results, including **a**, simulated off-axis hologram; **b**, corresponding SF domain; **c** and **d**, retrieved intensity distribution and phasefront of *H*-polarized component; **e** and **f**, retrieved intensity distribution and phasefront of *V*-polarized component; **g**, calculated polarization structure. **h-n,** Experimental results, including **h**, captured off-axis hologram; **i**, corresponding SF domain; **j** and **k**, retrieved intensity distribution and phasefront of *H*-polarized component; **l** and **m**, retrieved intensity distribution and phasefront of *V*-polarized component; **n**, calculated polarization structure.

Then we test the proposed detector by characterizing the vectorial structures of different vectorial structured light beams. Without loss of generality, we choose light beams on the HOPS of *l*=2 as an example. Such vectorial structured light beams can be mathematically expressed as,

$$\cos\theta OV_{+2}\widehat{R}+\sin\theta e^{i2\phi}OV_{-2}\widehat{L}, \tag{6}$$

where $2\theta$ and $2\phi$ are the polar angle and azimuthal angle on the HOPS of *l*=2, respectively. Here, $2\theta$ determines the power ratio between $OV_{+2}\widehat{R}$ and $OV_{-2}\widehat{L}$ states, while $2\phi$ serves as the relative phase between them. Varying $2\theta$ causes the light state to move along a meridian, whereas varying $2\phi$ leads to motion along a latitude. In this work, we present the characterization of the vectorial structures for two sets of vectorial structured light beams: one set lies on the equator, and the other lies on the meridian with a longitude of $2\phi= 0$, as displayed in Fig. 3a.

Figure 3b presents both simulation and experimental results for characterizing the polarization structures of vectorial structured light beams on the equator of the HOPS. Six states are tested: by fixing $2\theta=\frac{\pi}{2}$ and varying $2\phi$ from 0 to $\frac{5}{6}\pi$ in steps of $\frac{\pi}{6}$, these vectorial states are generated. The captured off-axis holograms and the retrieved polarization structures of the measured beams are shown in Fig. 3b. As the relative phase term $2\phi$ changes, the transverse polarization structures rotate clockwise. When plotting the polarization ellipses, linear polarizations are colored in blue, while left-elliptical/circular and right-elliptical/circular polarizations are colored red and green, respectively. To distinguish linear polarization from elliptical/circular polarization, we characterize the ellipticity of each polarization state and set an ellipticity threshold. Specifically, points whose polarization ellipses fall within the threshold are plotted in blue.

Figure 3c presents both simulation and experimental results for measuring the polarization structures of vectorial structured light beams on the meridian at a longitude of $2\phi= 0$. By fixing $2\phi= 0$ and varying $2\theta$ from 0 to $\pi$, we generate 11 vectorial states for testing. As the power ratio term $2\theta$ changes, the dominant polarization components within the transverse plane transition from right circular/elliptical to linear and then to left circular/elliptical.

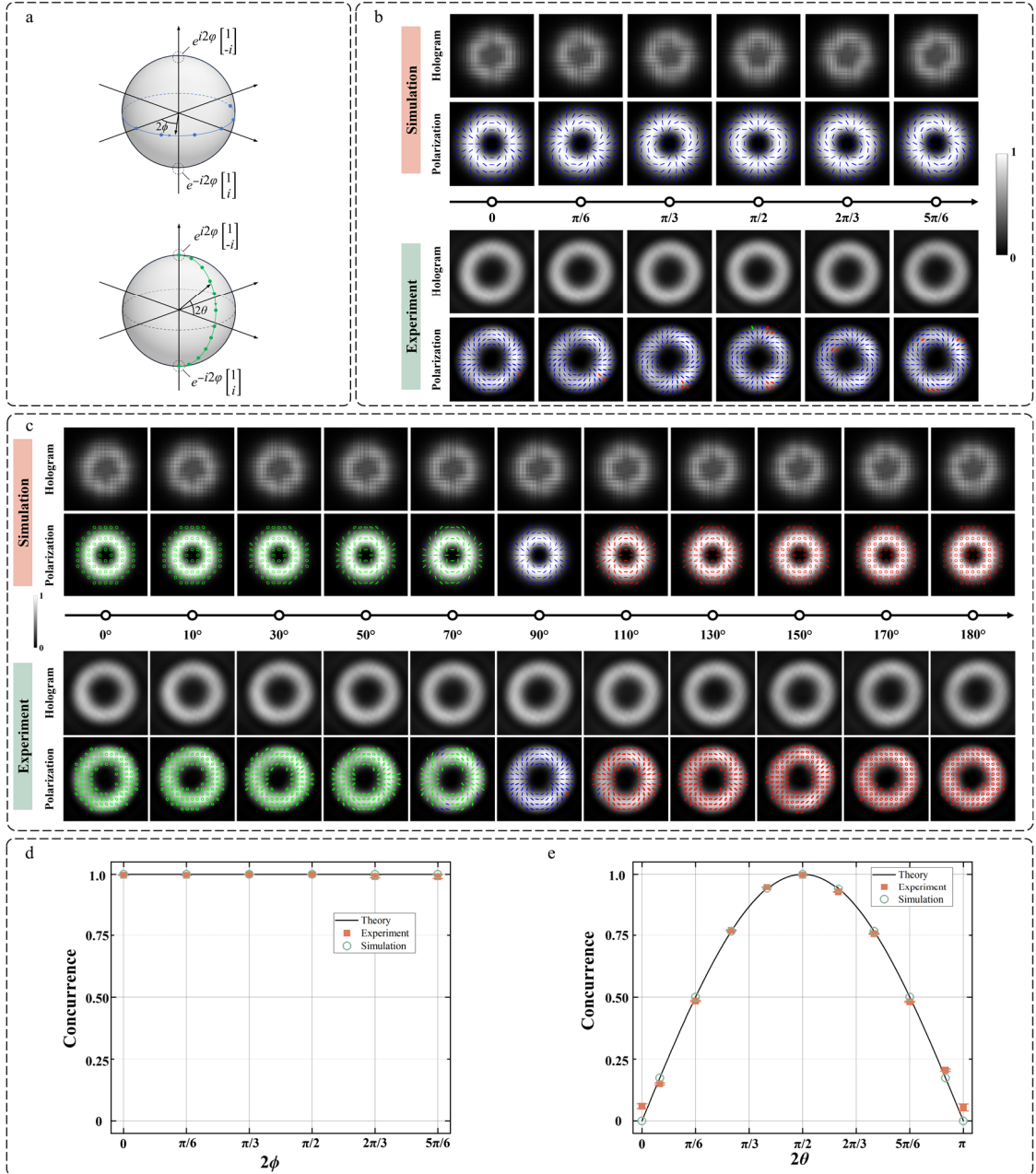

**Fig 3 | Simulation and experimental results for detecting the vectorial structures of different vectorial structured light beams on the HOPS with $l$=2. a,** Representation of all tested vectorial structured light beams on a HOPS with $l$=2. The top row corresponds to distinct vectorial structured light beams on the same equator, while the bottom row represents multiple vectorial structured light beams on the same meridian. **b,** Simulation and experimental results for detecting the vectorial structured light beams on the same equator, with $2\phi$ varies from 0 to $\frac{5}{6}\pi$ in steps of $\frac{\pi}{6}$. **c,** Simulation and experimental results for detecting the vectorial structured light beams on the same meridian, with $2\theta$ varies from -90° to 90°. **d,** Simulation and experimental results for characterizing the concurrence of generated vectorial structured light beams in **b**. **e,** Simulation and experimental results for characterizing the concurrence of generated vectorial structured light beams in **c**.

For vectorial light fields, an important metric describing the degree of entanglement is the concurrence (also referred to as vectorness, non-separability, or vector quality factor). The concurrence of vectorial light fields remains invariant in many complex media, making it a powerful metric for communication and sensing applications in such environments. Therefore, efficiently measuring the concurrence of different vectorial light fields is of great significance. Since our proposed detector retrieves the complex wavefronts in two orthogonal polarization channels, it provides all the necessary information to characterize concurrence values. We therefore apply our detector to characterize all the vectorial states generated in Fig. 3. For vectorial structured light beams expressed by Eq. (6), the concurrence equals $|\sin 2\theta|$. Consequently, for the vectorial structured light beams on the equator, all possess a concurrence of 1. Figure 3d shows the theoretical, simulated, and experimentally measured concurrence values for the six vectorial states on the equator of the HOPS. The black solid line represents the theoretical values, while hollow circles and solid squares denote the simulated and experimentally measured values, respectively. For the experimentally measured concurrence, 50 off-axis holograms were rapidly recorded, and the standard deviation of these 50 measurements is shown as the error bar in Fig. 3d. In addition, we characterize the concurrence values of 11 vectorial structured light states on the meridian at a longitude of $2\phi = 0$. As $2\theta$ varies from 0 to $\pi$ along this meridian, the concurrence first increases from 0 to 1 and then decreases back to 0. It is worth noting that low concurrence values correspond to cases where the power ratio between the two orthogonally polarized components is highly imbalanced. In such cases, imperfect polarization extinction of the experimental devices introduces larger measurement errors.

The above results demonstrate that the proposed detector is efficient in detecting the polarization structures and vectorial metrics of different vectorial light fields. Of course, the detector proposed in this paper still has some limitations in its use. Because the detector highly depends on multiplexing two orthogonal off-axis holograms in a single-shot process, the successful implementation of off-axis digital holography is the premise of the detector. To illustrate the boundaries of successful implementation of the proposed detector, we show some detection failure cases in Fig. 4.

For off-axis digital holography, the off-axis angle of the reference beam directly determines the spatial shift of the CC term within the spectral plane: a larger angle displaces the center of the CC term farther from the spectral origin. Thus, it is of great importance to choose suitable off-axis angles for the interference. For instance, the off-axis angle difference between two polarization channels should ensure a complete separation of $CC_1$ and $CC_2$ terms. As indicated in Fig. 4a, insufficient spectral separation between $CC_1$ and $CC_2$ terms results in incorrectly retrieved complex wavefronts. Furthermore, the off-axis angle between the tested beam and the reference beam should be large enough to ensure a complete separation of CC and DC terms. Shown in Fig. 4b is a case that breaks this requirement. In this case, the retrieved intensity profiles and phasefronts are all unwanted ones with high errors. The root cause of these distortions lies in the irreversible overlap of wanted spectral component and unwanted spectral components, which fundamentally compromises the fidelity of complex amplitude retrieval of off-axis digital holography. Thus, these requirements together put some lower limit of the off-axis angles of our proposed detector.

Nevertheless, the off-axis angle cannot be increased indefinitely. In our experiment, we systematically increase the off-axis angle and retrieved the intensity profiles of the tested beam. As shown in Fig. 4c, the absolute intensity of the demodulated signal decreases monotonically as the off-axis angle increases. This behavior arises from the changing microscopic structure of the interference fringes. A larger off-axis angle corresponds to a finer fringe period. When the fringe period approaches or falls below the pixel size of the detector, two adverse effects occur. First, a single pixel spatially averages over adjacent bright and dark fringes, reducing the fringe modulation depth and consequently the reconstructed intensity. Second, when the off-axis angle is too large, the carrier frequency may exceed the Nyquist limit of the detector, causing spectral aliasing. Aliasing manifests as pseudo-periodic fringes correlated with the pixel grid, leading to severe non-physical distortions in the reconstructed image. According to the Nyquist criterion, accurate recording of interference fringes without aliasing requires at least two pixels per fringe period.

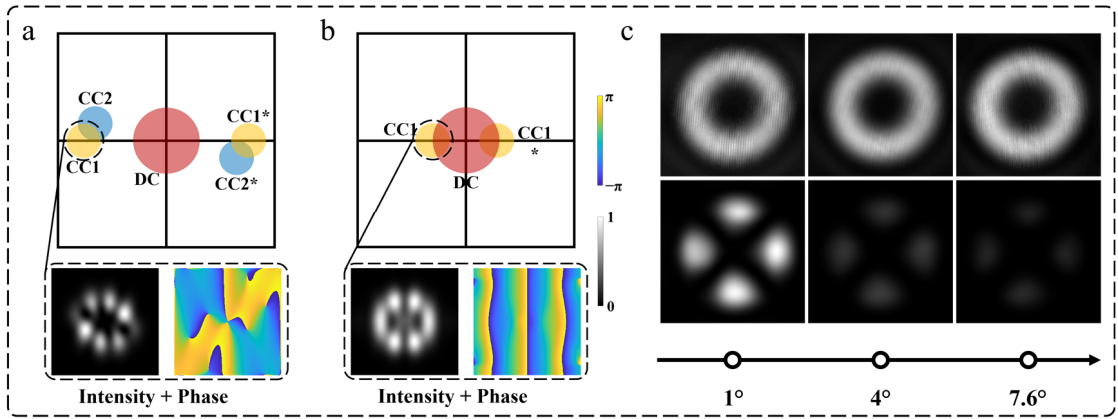

**Fig 4 | Some unsuitable conditions leading to detection failure — illustrating the boundaries of successful implementation of the proposed detector. a,** Incomplete separation of $CC_1$ and $CC_2$ terms. **b,** Incomplete separation of CC and DC terms. **c,** Excessively large off-axis angles introduced between the reference beam and the tested beam.

**Discussion**

In this work, we propose and experimentally demonstrate a single-shot holographic multiplexed detector capable of capturing the complete vectorial information of light fields. Unlike traditional Jones-matrix or Stokes-polarimetry methods that require sequential measurements or bulky polarization-beam-splitting optics in the signal path, our approach encodes the two orthogonal polarization channels into a single off-axis interference hologram using two reference beams with different carrier frequencies. A fast Fourier transform, followed by filtering and inverse transform, simultaneously recovers the complex wavefronts of both polarization components. The remaining global phase difference between the two reference arms is eliminated by a self-calibration procedure that relies on the equality of the mean phasefronts of the signal components—requiring no extra measurements. We have successfully applied the detector to a series of vectorial structured light beams on the HOPS of $l = 2$, retrieving their polarization structures and concurrence values in excellent agreement with simulations. We also analyze the operating conditions of the detector: the off-axis angles must be chosen to satisfy the requirements of off-axis digital holography.

Our approach offers several distinct advantages over existing techniques. For instance, the proposed detector provides a compact, efficient, and truly single-shot solution for spatially resolved vectorial light characterization. It eliminates temporal fluctuations and misalignment errors associated with sequential measurements, and the absence of polarization splitting devices in the signal path preserves the integrity of the test beam, which may be particularly beneficial for measuring weak or photosensitive samples. Nevertheless, the detector inherits the fundamental prerequisites of off-axis digital holography: a coherent light source is necessary to produce stable interference fringes, which places high demands on the coherence of the light source. Looking forward, we anticipate that the single-shot, full-field vectorial measurement capability will be particularly impactful in studies of light–matter interaction, real-time quantum state tomography, vectorial adaptive optics in complex media[39], and chiral sensing, as dynamical vectorial fields evolve on fast timescales and require real-time capture in these challenging scenarios.

**Methods**

**Concurrence measurement**

The concurrence value of vectorial light fields can be expressed as $\sqrt{1-s^2}$, where $s$ is the length of the Bloch vector, given by $s=\sqrt{\sum_{i=1}^{3}\langle\sigma_i\rangle^2}$. Here, $\sigma_1$, $\sigma_2$ and $\sigma_3$ represent the expectation values of the Pauli matrices. These expectation values can be calculated from the power coefficients of the field projected onto six different basis states, according to the following formula,

$$\begin{cases}\langle\sigma_1\rangle=\langle P_{3H}+P_{3V}\rangle-\langle P_{4H}+P_{4V}\rangle\\ \langle\sigma_2\rangle=\langle P_{5H}+P_{5V}\rangle-\langle P_{6H}+P_{6V}\rangle,\\ \langle\sigma_3\rangle=\langle P_{1H}+P_{1V}\rangle-\langle P_{2H}+P_{2V}\rangle\end{cases} \quad (7)$$

where $P_{iH}$ and $P_{iV}$ (with $i$=1,…,6) denote the power coefficients for the $H$- and $V$-polarized channels corresponding to the $i$-th basis state, respectively. These coefficients can be measured by digitally calculate the overlap integrals between measured wavefronts and six different basic states:[35]

$$\begin{cases}P_{iH}=\left|\iint E_H(x,y)E_{i,H}^*(x,y)\,dxdy\right|^2\\ P_{iV}=\left|\iint E_V(x,y)E_{i,V}^*(x,y)\,dxdy\right|^2\end{cases}, \quad (8)$$

where $E_H(x,y)$ and $E_V(x,y)$ are measured complex wavefronts in $H$- and $V$-polarized channels. $E_{i,H}^*(x,y)$ and $E_{i,V}^*(x,y)$ are the complex conjugations of the complex wavefronts for the $i$-th basic states in $H$- and $V$-polarized channels, respectively. For vectorial structured light beams expressed by Eq. (6), the six basic states include two pure OAM modes $OV_{+2}$ and $OV_{-2}$, and four superposition states $OV_{+2}+e^{i\beta}OV_{-2}$ characterized by different inter-modal phase $\beta$=0, $\frac{\pi}{2}$, $\pi$, $\frac{3\pi}{2}$.

**Data Availability Statement**

All data, theory details, simulation details that support the findings of this study are available from the corresponding authors upon reasonable request.

**Acknowledgments**

This work was supported by the National Natural Science Foundation of China (62405233), the

**Author contribution statement**

Y. L. developed the concept. Y. L. and H. X. conceived the experiment. S. Q. constructed the experiment, acquired the experimental data. All authors contributed to data analyses. Y. L., C. C., and X. Y. provided technical supports. Y. L. and H. X. finalized the paper. Y. L. and H. X. supervised the project.

**Conflicts of Interest**

The authors declare that there is no conflict of interest regarding the publication of this article.

# Supplementary information for: Seeing full vectorial structures of light fields with a single-shot holographic multiplexed detector

*Sitao Qin[+,1], Yize Liang[+,1,2]\*, Shuai Cao[1], Changqing Cao[1], Xukun Yin[1], Lixian Liu[1], Mingjian Cheng[3], and Huailiang Xu[1,4]\**

[1] School of Optoelectronic Engineering, Xidian University, Xi'an 710071, China

[2] Key Laboratory of Light Field Manipulation and System Integration Applications in Fujian Province, School of Physics and Information Engineering, Minnan Normal University, Zhangzhou 363000, China

[3] School of physics, Xidian University, Xi'an 710071, China

[4] State Key Laboratory of Integrated Optoelectronics, College of Electronic Science and Engineering, Jilin University, Changchun 130012, China

[+]These authors contributed equally to this work.

[*]E-mail: liangyize@xidian.edu.cn, xuhuailiang@xidian.edu.cn

**Supplementary Note 1. Simulation principle**

Here we introduce the principle for our numerical simulation process. Without loss of generality, we consider a vector beam constructed by the superposition of two optical vortex beams carrying opposite spin and orbital angular momentum (OAM). As illustrated in Fig. s1, two Laguerre-Gaussian (LG) beams with mode indices $l=\pm 2$ and radial index $p=0$ are employed. The scalar electric field of an $LG_{l,p}$ beam in cylindrical coordinates is described by:

$$LG_{l,p}(r,\varphi,z) = \left(\frac{2p!}{\pi(p+|l|)!}\right)^{1/2} \frac{1}{w(z)} \left[\frac{r\sqrt{2}}{w(z)}\right]^{|l|} \exp\left[-\frac{r^2}{w^2(z)}\right] L_p^{|l|}\left(\frac{2r^2}{w^2(z)}\right) \\ \exp\left[-i\frac{k_0 r^2 z}{2(z^2+z_R^2)}\right] \exp\left[-i(2p+|l|+l)\arctan\left(\frac{z}{z_R}\right)\right] \exp(il\varphi) \quad \text{(S1)}$$

where $(r,\varphi,z)$ are the cylindrical coordinates, $w(z)$ defines the $1/e$ radius of the beam at distance $z$, and $w(z) = w_0\left[(z^2+z_R^2)/z_R^2\right]^{1/2}$. Here $z_R$ is the Rayleigh range, $k_0 = \frac{2\pi}{\lambda}$ is the vacuum wave number, and $L_p^{|l|}(x)$ represents the associated Laguerre polynomial, related to the standard Laguerre polynomials $L_{p+|l|}$ by:

$$L_p^{|l|}(x) = (-1)^{|l|} \frac{d^{|l|}}{dx^{|l|}} L_{p+|l|}(x), \quad \text{(s2)}$$

In the simulation, to efficiently determine the Jones vectors of each pixel of light fields and plot the corresponding polarization structures, all light fields are simulated in the *H-* (horizontal) and *V-* (vertical) polarization channels. Thus, the *H*-polarization component $E_x$ and *V*-polarization component $E_y$ of the vector beam in Fig. s1 are given by:

$$\begin{cases} E_x = \cos\psi LG_{2,0}(r,\varphi,z) + \sin\psi LG_{-2,0}(r,\varphi,z)e^{i2\alpha} \\ E_y = \cos\psi LG_{2,0}(r,\varphi,z)e^{i\frac{\pi}{2}} + \sin\psi LG_{-2,0}(r,\varphi,z)e^{-i\frac{\pi}{2}}e^{i2\alpha} \end{cases} \quad \text{(s3)}$$

here, $\cos\theta$ and $\sin\theta$ determine the power ratio between $LG_{2,0}$ and $LG_{-2,0}$ components, while the factor $e^{i2\alpha}$ introduces a relative phase difference of $2\alpha$ between the two LG beams. The spherical angle parameters $(2\psi, 2\alpha)$ provide a geometric representation of all vectorial structured light beams on the higher-order Poincaré sphere (HOPS).

Subsequently, two off-axis Gaussian beams are generated to interfere with the vector beam. The two reference Gaussian beams can be described by:

$$\begin{cases} E_{Gaussian,x} = LG_{0,0}(r,\varphi,z)e^{-i\frac{2\pi x}{d_x}} \\ E_{Gaussian,y} = LG_{0,0}(r,\varphi,z)e^{-i\frac{2\pi y}{d_y}} \end{cases} \quad \text{(s4)}$$

where $E_{Gaussian,x}$ and $E_{Gaussian,y}$ are the scalar electric fields of *H-* and *V*-polarized reference Gaussian beams, $e^{-i\frac{2\pi x}{d_x}}$ and $e^{-i\frac{2\pi y}{d_y}}$ are $x$- and $y$-directional gratings which lead to $x$- and $y$-directional diffraction angles, $d_x$ and $d_y$ are the periods of such gratings. The resulting off-axis hologram formed by the three-beam interference can be expressed as:

$$I_{total}=|E_x+E_{Gaussian,x}|^2+|E_y+E_{Gaussian,y}|^2, \qquad (s5)$$

By performing a 2-dimensional(2D) FFT on the off-axis hologram, the SF domain is acquired. The CC terms are then filtered out, and applying a 2D IFFT to these terms enables the rapid retrieval of the complex wavefronts for the two orthogonal polarization components of the optical vortex beam. Subsequently, the polarization structure of the vector structured beam can be fully characterized by calculating the Jones vector at each pixel of the light field. In our simulation, the computational resolution of the light field is set to 300×300 pixels. To facilitate the overlay of the polarization map onto the intensity profile, the original field is divided into 15×15 superpixel regions. Each superpixel encompasses a 20×20 array of original pixels, and its effective Jones vector is determined by averaging the Jones vectors of all constituent 20×20 pixels. Plotting polarization ellipses on superpixel regions enables visualizing the polarization structure of simulated vector beam. In addition, the concurrence value of vector beams can also be characterized by calculating the overlapping degree between retrieved complex wavefronts and six basic spatial modes (see Supplementary Note 5).

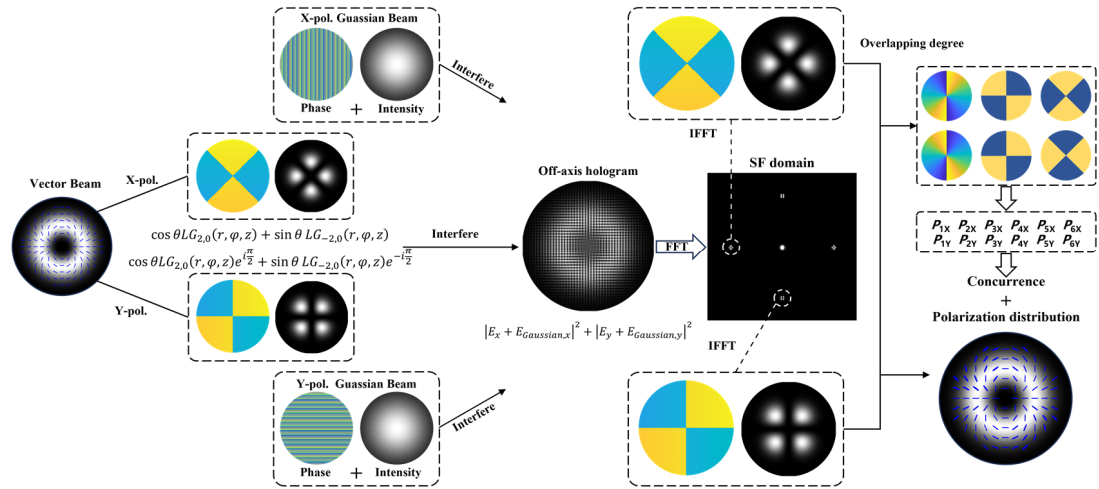

**Fig s1 | Simulation principle for single-shot characterizing the full vectorial structure of a vector beam.**

**Supplementary Note 2. Global phase calibration**

As mentioned in the Main Text, a global phase difference exists for two reference beams in the experiment. This phase discrepancy superimposes directly onto the phasefronts retrieved from the holograms, thereby introducing measurement errors on polarization structures. In this case, it is essential to calibrate such a global phase difference to accurately characterize the polarization structures of generated vectorial structured light beams.

Here we propose a self-calibrated approach to characterizing the global phase difference for calibration. For instance, considering the experimentally generated vector beam $|+2\rangle|L\rangle+|-2\rangle|R\rangle$, where $|+2\rangle$ and $|-2\rangle$ are OAM beams with topological charges of +2 and -2, $|L\rangle$ and $|R\rangle$ denote left and right circular polarization states. In the experiment, the measured phasefronts in $H$- and $V$- polarization channels can be described as:

$$\begin{cases} \varphi_{measured,H}(x,y)=\varphi_{S,H}(x,y)-\varphi_{R,H} \\ \varphi_{measured,V}(x,y)=\varphi_{S,V}(x,y)-\varphi_{R,V} \end{cases}, \tag{s6}$$

where $\varphi_{measured,H}(x,y)$ and $\varphi_{measured,V}(x,y)$ are measured phasefronts in $H$- and $V$- polarization channels, respectively. The measured phasefronts include the accurate phasefronts of signal beams — i.e., the scalar phasefronts of the $H$-and V-polarized components of the measured vector beam, denoted as $\varphi_{S,H}(x,y)$ and $\varphi_{S,V}(x,y)$ — as well as the global phase terms of the reference beams, $\varphi_{R,H}$ and $\varphi_{R,H}$. Here, for better clarification, we illustrate these phase terms in Fig. s2. Note that the integral of $\varphi_{S,H}(x,y)$ and $\varphi_{S,V}(x,y)$ over the whole plane is always the same, we have:

$$\frac{1}{A}\iint_A \varphi_{S,H}(x,y)dxdy = \frac{1}{A}\iint_A \varphi_{S,V}(x,y)dxdy \tag{s7}$$

where $A$ denotes the transverse beam plane. That is, the mean value of $\varphi_{S,H}(x,y)$ and $\varphi_{S,V}(x,y)$ are the same. Considering calculating the mean value of $\varphi_{measured,H}(x,y)$ and $\varphi_{measured,V}(x,y)$, which can be written as:

$$\overline{\varphi_{measured,H}(x,y)}-\overline{\varphi_{measured,V}(x,y)}=\left(\overline{\varphi_{S,H}(x,y)}-\varphi_{R,H}\right)-\left(\overline{\varphi_{S,y}(x,y)}-\varphi_{R,V}\right)=\varphi_{R,V}-\varphi_{R,H}, \tag{s8}$$

where $\overline{\varphi_{measured,H}(x,y)}$ and $\overline{\varphi_{measured,V}(x,y)}$ denote the mean value of $\varphi_{measured,H}(x,y)$ and $\varphi_{measured,V}(x,y)$ respectively, $\overline{\varphi_{S,H}(x,y)}$ and $\overline{\varphi_{S,V}(x,y)}$ are the mean value of $\varphi_{S,H}(x,y)$ and $\varphi_{S,V}(x,y)$ respectively. Due to $\varphi_{R,H}$ and $\varphi_{R,V}$ are global phase term, their mean value are still themselves. Thus, we finally obtain the global phase difference for calibration.

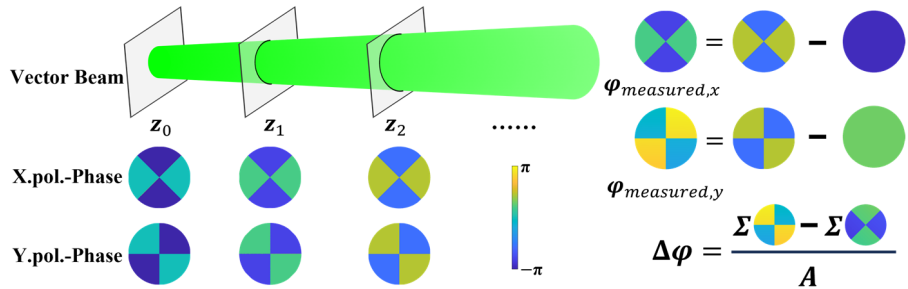

**Fig s2** | Illustration of performing phase calibration process.

It is worth mentioning that such a digital self-calibrated approach is efficient for accurately measure the polarization structures of vectorial structured light beams, which doesn't require any additional measurements. This approach can also be extended to other situations, as long as there is some connection between the mean values of $\varphi_{S,H}(x,y)$ and $\varphi_{S,V}(x,y)$. Obviously, there maybe some extreme cases where $\varphi_{S,H}(x,y)$ and $\varphi_{S,V}(x,y)$ are both random phasefronts. In such cases, one can adopt the configuration illustrated in Fig. S2 for pre-calibration; that is, only one additional measurement is needed to obtain the calibrated phase, thereby providing a benchmark for a series of subsequent measurements of different random vectorial light fields.

**Supplementary Note 3. Experimental setup**

Figure s3 illustrates the experimental setup for our proposed single-shot holographic multiplexed vectorial structured light detector. As indicated in Fig. s3, a 532-nm laser creates a free-space Gaussian beam. After passing through a polarizer (Pol. 1), the Gaussian beam is converted into a linearly polarized Gaussian beam. Then a beam splitter separates the Gaussian beam into two parts, one is the signal arm (direct path), and the other is the reference arm (reflective path).

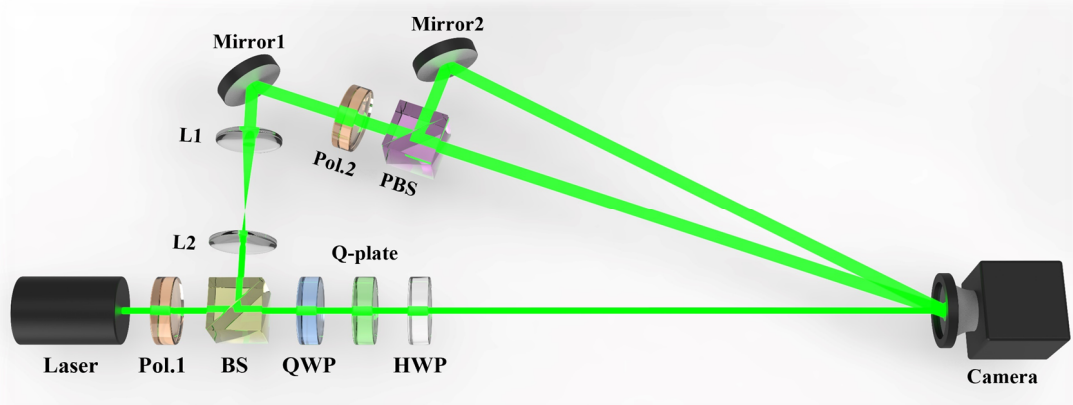

**Fig s3 | Experimental setup for the single-shot holographic multiplexed detector.** Pol., polarizer; BS, beam splitter; QWP, quarter-wave plate; HWP, half-wave plate; L, lens; PBS, polarization beam splitter.

For the signal arm, a Q-plate is utilized to convert an input Gaussian beam into a vectorial structured light beam. A quarter-wave plate is applied to create different kinds of vectorial structured light beams which can be represented as distinct points on the meridians of a higher-order Poincaré sphere (HOPS). Similarly, a half-wave plate (HWP) is used to provide phase difference between two circularly polarized optical vortex beams. Thus, the HWP is applied to produce variations on the latitudes of a HOPS (see Supplementary Note 4 for more details).

For the reference arm, a 3× beam expander consists of two lenses (L1, *f*=5 cm, and L2, *f*=15 cm) is utilized to enlarge the Gaussian beam to generate a large beam as a reference beam for off-axis digital holographic multiplexing. Pol. 2 ensures the diagonal polarization of the reference beam, so that it can be divided in to two parts by a polarization beam splitter (PBS). And the two parts possess the same power yet orthogonal polarizations (H and V polarizations). Finally, both the two reference beams off-axis interfere with the signal beam with different off-axis angles. A camera captures the multiplexed off-axis interference hologram to achieve single-shot vectorial structure reconstruction.

**Supplementary Note 4. Control the states of measured vectorial structured light fields**

This section details the method for controlling the states of the measured vectorial structured light. We use a Q-plate with *q*=1 to generate vectorial structured light beams with a modal index of 2. Upon a Gaussian beam incident on the Q-plate, opposite SAM states experience opposite OAM modulations, as expressed by,

$$\begin{cases} |0\rangle|R\rangle \to |+2q\rangle|L\rangle \\ |0\rangle|L\rangle \to |-2q\rangle|R\rangle \end{cases}, \tag{S9}$$

where $|0\rangle$ denotes the fundamental Gaussian mode, $|+2q\rangle$ and $|-2q\rangle$ are OAM modes with topological charges of $2q$ and $-2q$, $|L\rangle$ and $|R\rangle$ correspond to left circular polarization and right circular polarization respectively. Thus, a linear polarization state input results in a full vector beam output, that is, vectorial structured light on the equator of +2 HOPS. In the experiment, two series of vectorial structured light beams on the +2 HOPS are generated to test the efficiency of the proposed detector, namely states on the equator or meridian.

To generate distinct states on the equator, an adjustable relative phase is required when forming a 1:1 superposition of the $|+2q\rangle|L\rangle$ and $|-2q\rangle|R\rangle$ states. This is achieved using the HWP shown in Fig. S3 in the Supplementary Information. Considering the Jones matrix of a HWP as $\begin{bmatrix} i\cos2\theta & i\sin2\theta \\ i\sin2\theta & -i\cos2\theta \end{bmatrix}$, the transmission of the $|L\rangle$ or $|R\rangle$ state through it can be described as follows,

$$|L\rangle: \begin{bmatrix} \cos2\theta & \sin2\theta \\ \sin2\theta & -\cos2\theta \end{bmatrix}\begin{bmatrix} 1 \\ i \end{bmatrix} = \begin{bmatrix} \cos2\theta + i\sin2\theta \\ \sin2\theta - i\cos2\theta \end{bmatrix} = \begin{bmatrix} 1 \\ -i \end{bmatrix} e^{i2\theta}, \tag{S10}$$

$$|R\rangle: \begin{bmatrix} \cos2\theta & \sin2\theta \\ \sin2\theta & -\cos2\theta \end{bmatrix}\begin{bmatrix} 1 \\ -i \end{bmatrix} = \begin{bmatrix} \cos2\theta - i\sin2\theta \\ \sin2\theta + i\cos2\theta \end{bmatrix} = \begin{bmatrix} 1 \\ i \end{bmatrix} e^{-i2\theta}, \tag{S11}$$

where $\theta$ suggests that the angle between the $x$ axis and HWP's fast axis. Thus, adjusting the HWP leads to a relative global phase of $4\theta$, which creates multiple vectorial structured light states on the same equator.

To generate different states on the same meridian, one needs to control the power ratio of two $|+2q\rangle|L\rangle$ and $|-2q\rangle|R\rangle$ components. This is accomplished using the QWP illustrated in Fig. S3 in the Supplementary Information. Considering a linear polarization state $\begin{bmatrix} A \\ B \end{bmatrix}$ input the QWP, such a Jones matrix can be projected onto the fast and slow axes, as shown in Fig. S4. The newly obtained matrix is a polarization matrix with respect to the fast and slow axes, given by:

$$\begin{cases} A'=A\cos\alpha + B\sin\alpha \\ B'=A\sin\alpha - B\cos\alpha \end{cases}, \tag{S12}$$

where $\begin{bmatrix} A' \\ B' \end{bmatrix}$ represents the newly obtained polarization matrix, and $\alpha$ is the angle between the QWP's fast axis and the $x$ axis. The resulting vector in the fast–slow basis is given by:

$$\begin{bmatrix} A' \\ B' \end{bmatrix} = \begin{bmatrix} \cos\alpha & \sin\alpha \\ \sin\alpha & -\cos\alpha \end{bmatrix} \begin{bmatrix} A \\ B \end{bmatrix}, \tag{S13}$$

Next, the effect of the QWP is taken into account. The QWP introduces a phase delay of $\frac{\pi}{2}$ between the fast and slow axes, which can be written as:

$$\begin{bmatrix} A'' \\ B'' \end{bmatrix} = \begin{bmatrix} 1 & 0 \\ 0 & e^{i\frac{\pi}{2}} \end{bmatrix} \begin{bmatrix} A' \\ B' \end{bmatrix}, \tag{S14}$$

We then perform the inverse coordinate transformation (from the fast/slow axes back to the $x/y$ axes) on $\begin{bmatrix} A'' \\ B'' \end{bmatrix}$ to obtain the output Jones vector. Thus, the final Jones vector describing the polarization state of the light field exiting the QWP is:

$$\begin{bmatrix} C \\ D \end{bmatrix} = \begin{bmatrix} \cos\alpha & \sin\alpha \\ \sin\alpha & -\cos\alpha \end{bmatrix} \begin{bmatrix} 1 & 0 \\ 0 & e^{i\frac{\pi}{2}} \end{bmatrix} \begin{bmatrix} \cos\alpha & \sin\alpha \\ \sin\alpha & -\cos\alpha \end{bmatrix} \begin{bmatrix} A \\ B \end{bmatrix}, \tag{S15}$$

where $\begin{bmatrix} C \\ D \end{bmatrix}$ is the output Jones matrix, and $\alpha$ is the angle between the fast axis and $x$ axis of QWP.

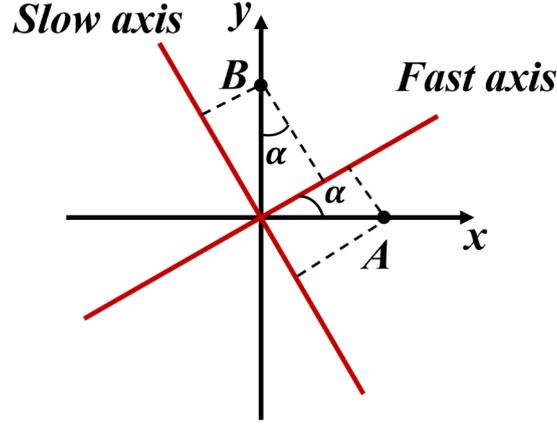

**Fig s4** | Illustration of the coordinate transformation process of a QWP.

Note that there is a polarizer before the QWP (Pol. 1), the polarization state input the QWP is controlled to a horizontal polarization state with a Jones matrix of $\begin{bmatrix}1\\0\end{bmatrix}$. Hence, Eq. (S15) can be expressed as,

$$\begin{bmatrix}C\\D\end{bmatrix}=\begin{bmatrix}\cos\alpha & \sin\alpha\\ \sin\alpha & -\cos\alpha\end{bmatrix}\begin{bmatrix}1 & 0\\ 0 & e^{i\frac{\pi}{2}}\end{bmatrix}\begin{bmatrix}\cos\alpha & \sin\alpha\\ \sin\alpha & -\cos\alpha\end{bmatrix}\begin{bmatrix}1\\0\end{bmatrix}$$

$$=\begin{bmatrix}\cos^2\alpha+i\sin\alpha\\ \cos\alpha\sin\alpha-i\cos\alpha\sin\alpha\end{bmatrix}$$

$$=\sqrt{2}\cos\left(\frac{\pi}{4}-\alpha\right)e^{i\alpha}\begin{bmatrix}1\\i\end{bmatrix}+\sqrt{2}\sin\left(\frac{\pi}{4}-\alpha\right)e^{-i\alpha}\begin{bmatrix}1\\-i\end{bmatrix} \quad (S16)$$

This means that by adjusting the QWP to control the parameter $\alpha$, one can obtain a superposition of two circularly polarized beams with different power ratios. Consequently, the power ratio of the two components $|+2q\rangle|L\rangle$ and $|-2q\rangle|R\rangle$ can be tailored to generate different states on the same meridian of a HOPS.

**Supplementary Note 5. Additional results for vectorial structured light measurements in the Main Text**

In this section, we show some additional results for vectorial structured light measurements in the Main Text. Displayed in Fig. s5 are additional results for measuring the full vectorial structures of vectorial structured light beams on the equator of HOPS, corresponding to results in Fig. 3 b in the main text. We detail the transferred spatial frequency (SF) domain, retrieved *H*- and *V*- polarized intensity profiles and phasefronts of the measured multiplexed hologram in Fig. 3 b in the main text. Both simulation and experimental results are shown in Fig. s5. Based on these detailed results, one can characterize the full vectorial structures of generated vectorial structured light beams.

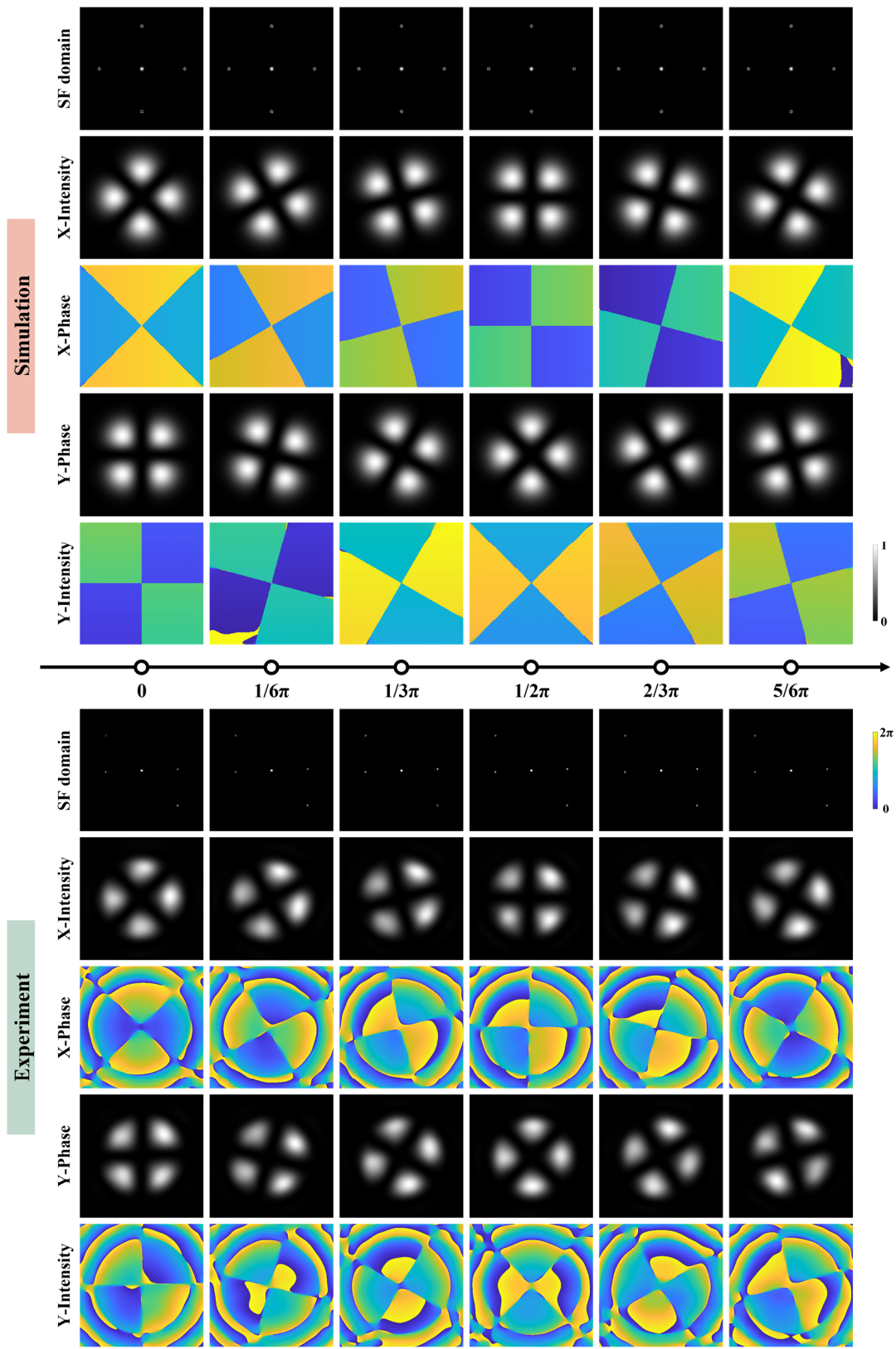

**Fig s5 | Additional experimental results for measuring the full vectorial structures of vectorial structured light beams on the equator of HOPS, which corresponds to results in Fig. 3 b in the main text.**

Figure S6 presents additional results for measuring the full vectorial structures of vectorial structured light beams on the meridian of the HOPS, corresponding to the results shown in Fig. 3c of the main text. Similarly, we detail the transferred spatial frequency (SF) domain, along with the retrieved *H*- and *V*-polarized intensity profiles and phasefronts of the measured multiplexed hologram from Fig. 3c. Both simulation and experimental results are displayed in Fig. S6. Based on these detailed results, one can fully characterize the vectorial structures of the generated vectorial structured light beams.

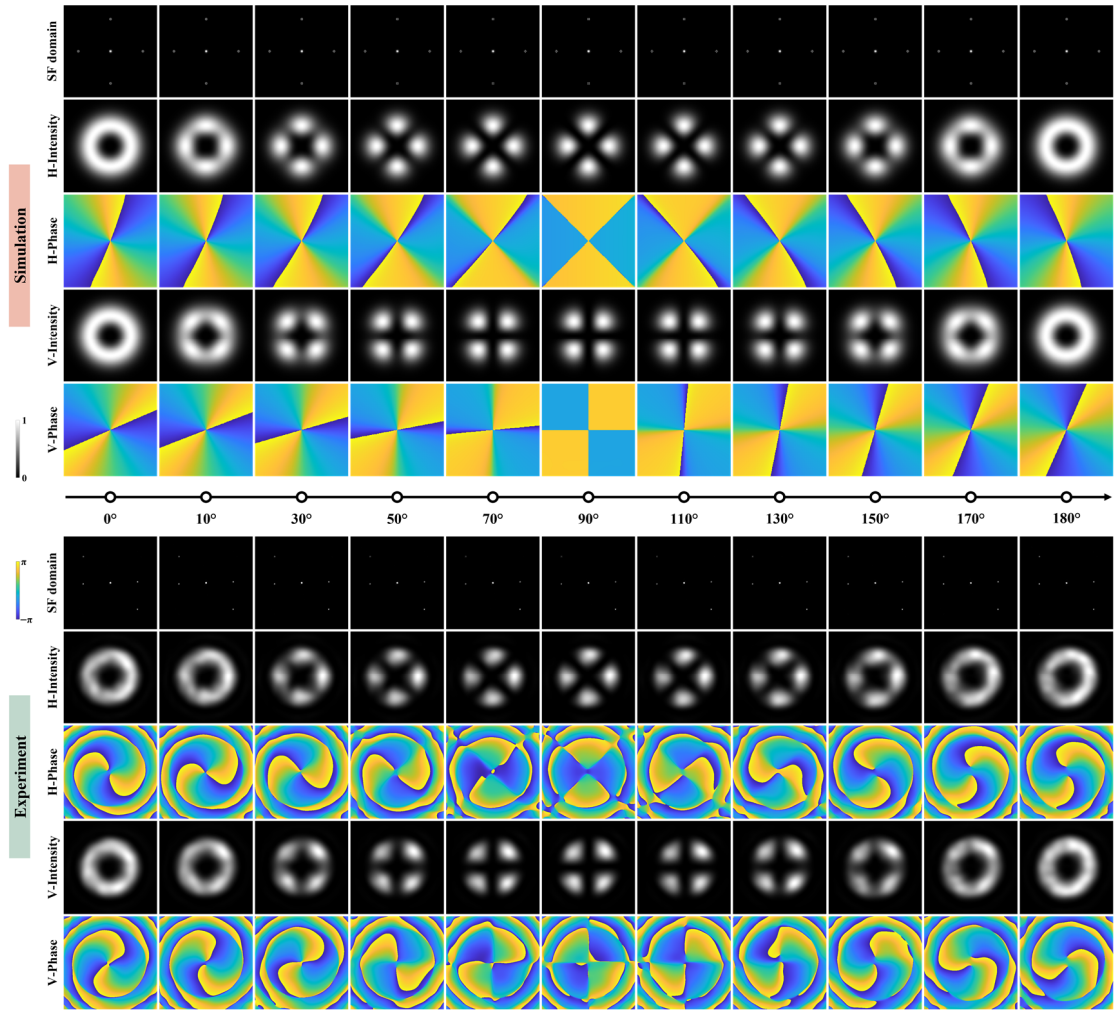

**Fig s6 | Additional experimental results for measuring the full vectorial structures of vectorial structured light beams on the meridian of HOPS, which corresponds to results in Fig. 3 c in the main text.**